\documentclass[12pt]{article}

\usepackage{graphicx}

\def\um{{1}/{2}}
\def\sq{1/\sqrt{2}}

\newcommand{\lsim}{ \mathop{}_{\textstyle \sim}^{\textstyle <} }
\begin{document}
\renewcommand{\thefootnote}{\fnsymbol{footnote}}
\renewcommand{\theenumi}{(\roman{enumi})}
\title{Cosmological Constraint on the Zee model}
\author{
{N. Haba$^{1,2}$}\thanks{haba@eken.phys.nagoya-u.ac.jp}
{, K. Hamaguchi$^3$}\thanks{hama@hep-th.phys.s.u-tokyo.ac.jp}
{, and Tomoharu Suzuki$^2$}\thanks{tomoharu@eken.phys.nagoya-u.ac.jp}
\\
\\
\\
{\small \it $^1$Faculty of Engineering, Mie University,}
{\small \it Tsu Mie 514-8507, Japan}\\
{\small \it $^2$Department of Physics, Nagoya University,}
{\small \it Nagoya, 464-8602, Japan}\\
{\small \it $^3$Department of Physics, University of Tokyo, }
{\small \it Tokyo, 113-0033, Japan}}

\maketitle
\date{}

\vspace{-10.5cm}
\begin{flushright}
hep-ph/0108013\\
DPNU-01-18\\
UT-956\\
\end{flushright}
\vspace{10.5cm}
\vspace{-2.5cm}
\begin{center}
\end{center}
\renewcommand{\thefootnote}{\fnsymbol{footnote}}
%
%
\begin{abstract}
It is well known that the Zee model induces small neutrino masses by
 radiative corrections, where the bi-maximal flavor mixing is
 possible. We analyze the cosmological condition in order for the baryon
 asymmetry generated in the early universe not to be washed out in the
 Zee model. Since the lepton number is violated explicitly in the Zee
 model, the baryon asymmetry might be washed out through the sphaleron
 processes together with the lepton-number violating interactions.  In
 this letter, we will show that the baryon asymmetry is {\it not} washed
 out, although it has been said that the Zee model cannot preserve the
 baryon asymmetry generated in the early universe.  This can be seen by
 considering an approximately conserved number, $L' \equiv
 L_e-L_\mu-L_\tau$.
\end{abstract}

\newpage
\section{Introduction}

Recently the evidences of neutrino oscillations are strongly supported
by both of the atmospheric~\cite{Kamiokande,SKatm} and the solar
neutrino
experiments~\cite{SKsolar,SNO,Cl-Homestake,Ga-Gallex-and-GNO}. The
former suggests an almost maximal lepton flavor mixing between the 2nd
and the 3rd generations, while the favorable solution to the solar
neutrino deficits is given by large mixing angle solution between the
1st and the 2nd generations ( LMA, LOW or VO )~\cite{post-SNO-analysis}.
Neutrino oscillation experiments indicate that the neutrinos have tiny
but finite masses, with two mass squared differences $\Delta m_{\odot}^2
< \Delta m_{\rm atm}^2$.  Since the neutrino mass is much smaller than
other quarks and leptons, the origin of the neutrino mass is expected to
be different from that of others.  The most popular mechanism to induce
the small neutrino masses is the seesaw mechanism~\cite{seesaw}, which
needs heavy right-handed neutrinos.  However, it is important to
consider also other possible scenarios which can explain the small
neutrino masses, especially low-energy extensions of the standard model.
The Zee model~\cite{zee} is such an alternative, which induces small
neutrino masses by radiative corrections, where the lepton number is
violated explicitly.

Generally, if there are lepton number violating interactions, care has
 to be taken for these interactions not to be too strong at the
 temperatures above the electroweak scale, since the baryon asymmetry
 generated in the early universe might be washed out by the
 ``sphaleron'' process~\cite{sphaleron}, which violates a linear
 combination of baryon ($B$) and lepton ($L$) number, $B+L$.  If the
 rate of the lepton number violating process becomes faster than the
 Hubble expansion rate $H$ during the epoch when the sphaleron process
 is in thermal equilibrium, the baryon asymmetry generated at higher
 temperature would be washed out, and there would be no matter
 anti-matter asymmetry in the present universe,\footnote{In this letter
 we assume that the production of the baryon asymmetry took place in the
 early universe before the electroweak phase transition.} which
 conflicts with the observation.

In this letter, we analyze the cosmological condition in order for the
 baryon asymmetry generated in the early universe not to be washed out
 in the Zee model. (We will not discuss the production mechanism of the
 baryon asymmetry. Instead, we just assume that the desired amount of
 the baryon asymmetry is generated in the early universe and discuss
 whether or not this baryon asymmetry can be preserved against the
 lepton number violating interaction in the Zee model together with the
 sphaleron process.)  In most part of the analysis, we take the LOW
 solution to the solar neutrino problem. (We shall note on the cases of
 other solutions in the last section.) We find that the baryon asymmetry
 is {\it not} washed out, although it has been said that the Zee model
 cannot preserve the baryon asymmetry~\cite{Sarkar}. This can be easily
 shown from the view point of a new lepton number $L' \equiv
 L_e-L_\mu-L_\tau$.

\vspace{5mm}
\section{Brief review of the Zee model}

Let us give a brief overview of the Zee
 model~\cite{zee,bimaximal,ST,noLMA,Higgs-pheno,Zee-recent} at
 first. The Zee model is a simple extension of the standard model, which
 has two Higgs doublets $\phi_i = (\phi_i^0, \phi_i^-)^T$ ($i=1,2$) and
 one $SU(2)_L$-singlet charged Higgs field (Zee singlet) $\omega^{\pm}$.
 The Zee model has the following interactions in addition to the
 standard model ones;
\begin{eqnarray}
 \Delta{\cal L}
  =
  f_{\alpha\beta}
  \,
  \overline{l_{\alpha_L}}^C
  (i\tau_2)
  l_{\beta_L}
  \,
  \omega^+
  +
  \mu \phi_2^T (i \tau_2) \phi_1\, \omega^+  
  +
  h.c.
  \label{Zee_Lag}\,,
\end{eqnarray}
where $l_{\alpha_L}$ denote the left-handed lepton doublets with flavor
indices $\alpha,\beta=e,\mu,\tau$.  Notice that the coupling $f_{\alpha
\beta}$ is anti-symmetric for the flavor index.  For simplicity, we have
omitted the Higgs potential $V(\phi_1,\phi_2,\omega^-)$ which are
irrelevant to the following discussion.  As for the Yukawa interaction,
we assume only $\phi_1$ couples to lepton fields as,
\begin{eqnarray}
 {\cal L}_{\rm Y}  &=&  
  +\overline{e_{\alpha_R}}
  \left(y_e\right)_{\alpha}
  \widetilde{\phi_1}^{\dagger }l_{\alpha_L}
  +h.c.
  \label{L-Yukawa} \; ,
\end{eqnarray}
where $\widetilde{\phi_1}\equiv (i \tau_2)\phi_1^*$.  As can be seen
from Eqs.~(\ref{Zee_Lag}) and (\ref{L-Yukawa}), the lepton number $L$ is
explicitly violated in the Zee model.\footnote{We assume that there is a
mixing term between $\phi_1$ and $\phi_2$\,, $m_3^2 \phi_1^{\dagger}
\phi_2 + h.c.$\,, in $V(\phi_1,\phi_2,\omega)$.}

In the charged Higgs sector, the Zee singlet $\omega$ is mixed with the
charged Higgs boson $\Phi^-$ through the $\mu\,\phi_1 \phi_2 \omega$
interaction in Eq.~(\ref{Zee_Lag});
\begin{eqnarray}
 \Phi^-&=&\cos \chi \;S_{1}^- -\sin \chi \;S_2^- 
  \,,
  \nonumber\\
 \omega^-  &=&\sin \chi \;S_1^- +\cos \chi \;S_2^- 
  \label{chi}  \,.
\end{eqnarray}
Here, $\Phi^- = \cos\beta \; \phi_1^- - \sin\beta \; \phi_2^-$ is the
charged Higgs boson which is orthogonal to the would-be Goldstone boson
after the neutral Higgs fields acquire vacuum-expectation values (VEVs)
$\langle \phi_i^{\;0} \rangle = v_i$, where $\tan \beta \equiv \langle
\phi_1^{\;0} \rangle$/$\langle \phi_2^{\;0} \rangle =v_1/v_2$.  We
denote the mass eigenstates of the charged Higgs sector as $S_i^{\pm}$
and their mass $m_{S_i}$.

\begin{figure}
 \setlength{\unitlength}{1cm}
 \begin{center}
  \begin{picture}(5,5)
   \put(-2,0){\includegraphics{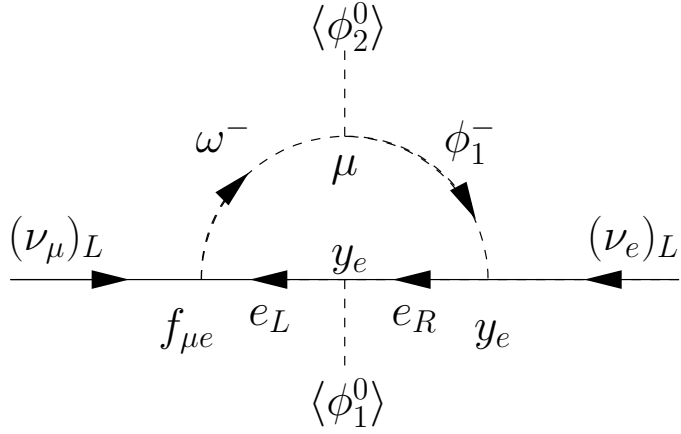}}
   \put(-2,1.7){\Large $(\nu_\mu)_L$}
   \put(0,.5){\Large $f_{\mu e}$}
   \put(2,-.5){\Large $\langle \phi_1^0 \rangle$}
   \put(1.2,.7){\Large $e_L$}
   \put(3.1,.7){\Large $e_R$}
   \put(4.2,.5){\Large $y_e$}
   \put(5.7,1.7){\Large $(\nu_e)_L$}
   \put(2.3,1.5){\Large $y_e$}
   \put(2,4.5){\Large $\langle \phi_2^0 \rangle$}
   \put(.5,3){\Large $\omega^-$}
   \put(3.8,3){\Large $\phi_1^-$}
   \put(2.3,2.7){\Large $\mu$}
  \end{picture}
 \end{center}
 \caption{Neutrino mass diagram in the Zee model}
 \label{fig:Zee_mass}
\end{figure}

In the Zee model, neutrino masses are generated by radiative
 corrections, as shown in Fig.~\ref{fig:Zee_mass}, and hence this model
 could provide an explanation of the smallness of neutrino masses.  The
 element of the mass matrix, generated by radiative correction at one
 loop level, is given by
\begin{eqnarray}
 m_{\alpha \beta}
  =
  f_{\alpha \beta}
  (m_{l_\beta}^2 - m_{l_\alpha}^2)
  \mu 
  \cot\beta
  \frac{1}{16 \pi^2}
  \frac{1}{m_{S_1}^2 - m_{S_2}^2} 
  \ln 
  \frac{m_{S_1}^2}{m_{S_2}^2} 
  \label{Zee_mass}\;.
\end{eqnarray}
Here $m_{l_\alpha}(\alpha=e,\mu,\tau)$ are the charged lepton masses.
In Eq.~(\ref{Zee_mass}), we have used $m_{S_i}\gg m_{l_\alpha}$.

Since the coupling constants $f_{\alpha \beta}$ are antisymmetric for
 the indices $\alpha,\beta$, the mass matrix Eq.~(\ref{Zee_mass}) is
 traceless.\footnote{ If we consider higher order loops, non-zero values
 appear in the diagonal elements. We shall neglect them, since the
 conclusion does not change.} The above Zee mass matrix has been
 analyzed~\cite{bimaximal} in the light of recent neutrino-oscillation
 experiments, and it was shown that it must be the following bi-maximal
 form to explain the experimental results;
\begin{eqnarray}
 M_\nu 
  &=&\left(
      \begin{array}{ccc}
       0 & m_{e\mu}& m_{e\tau} \\
       m_{e\mu}& 0 & m_{\mu\tau} \\
       m_{e\tau}& m_{\mu\tau}& 0
      \end{array}
      \right)
   \quad \simeq \quad
   m_0\left(
       \begin{array}{ccc}
	0&-\sq&\sq\\
	-\sq&0&\epsilon\\
	\sq&\epsilon&0
       \end{array}
       \right)
    \label{Zee_mass_matrix}     \,,
\end{eqnarray}
where $m_0^2=m_{e\mu}^2+m_{e\tau}^2$ , $\epsilon ={m_{\mu\tau}}/{m_0}$ ,
$m_{e\mu}\simeq -m_{e\tau}$ and $\epsilon \ll 1$.  The eigenvalues and
the MNS matrix induced from the neutrino mass matrix
Eq.~(\ref{Zee_mass_matrix}) is given by
\begin{eqnarray}
 m_1 \simeq  m_0(1 - \um\epsilon)
  \qquad m_2 \simeq - m_0(1+ \um\epsilon),  
  \qquad m_3 \simeq  m_0 \; \epsilon
  \,,
\end{eqnarray}
and
\begin{eqnarray}
 U_{\mathrm{MNS}}& \sim &
  \left(
   \begin{array}{ccc}
    \sq&\sq&0\\
    -\um&\um&\sq\\
    \um&-\um&\sq
   \end{array}
   \right)
   \,.
\end{eqnarray}
The neutrino mass hierarchy is shown in Fig.~\ref{fig:mass}. The mixing
angle between the 1st and the 2nd generations is maximal as well as the
mixing angle between the 2nd and the 3rd generations ($\theta_{12}\sim
\theta_{23}\sim 45^\circ$).  {}From oscillation experiments, the
eigenvalues of the mass matrix have to satisfy the following relations,
\begin{eqnarray}
 m_1^2-m_3^2 
  \simeq m_2^2-m_3^2  
  &\simeq& 
  m_0^2 \nonumber\\
 &\simeq& 2 m_{e\mu}^2 \simeq 2 m_{e\tau}^2 \simeq \Delta m_{\rm atm}^2
  \label{mass_atm}\;, \\
 m_2^2-m_1^2 
  &\simeq& 2 m_0^2  \; \epsilon 
  \nonumber \\
 &\simeq&
  2 m_0 m_{\mu\tau} \simeq \Delta m_{\odot}^2
  \label{mass_sol}
  \,.
\end{eqnarray}
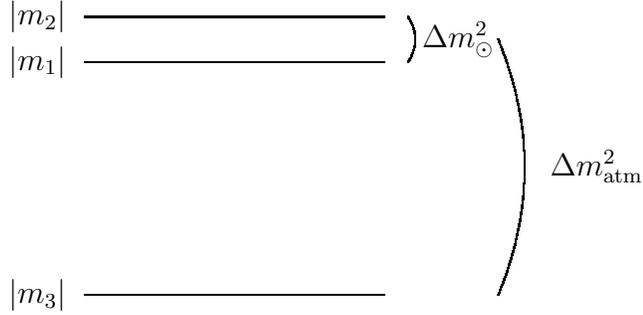
\begin{figure}
 \begin{center}
  \setlength{\unitlength}{1cm}
  \begin{picture}(6,5)
   \put(0,0){$|m_3|$}
   \put(1,.1){\line(1,0){4}}
   
   \put(0,3.1){$|m_1|$}
   \put(1,3.2){\line(1,0){4}}
   
   \put(0,3.7){$|m_2|$}
   \put(1,3.8){\line(1,0){4}}
   
   \qbezier(5.3,3.2)(5.5,3.5)(5.3,3.8)
   
   \qbezier(6.5,3.5)(7.2,1.8)(6.5,.1)
   
   \put(5.5,3.4){$\Delta m_{\odot}^2$}
   \put(7.2,1.7){$\Delta m_{\rm atm}^2$}
  \end{picture}
 \end{center}
 \caption{The hierarchy of the neutrino masses in the Zee model.}
 \label{fig:mass}
\end{figure}
Then, the relations among $f_{e\mu}$ , $f_{e\tau}$ and
 $f_{\mu\tau}$ should be  
 \begin{eqnarray}
  \left|\frac{f_{e\mu}}{f_{e\tau}}\right|
   &\simeq& \frac{m_{\tau}^2}{m_{\mu}^2}\simeq 3\times10^2\; ,\\
  \left|\frac{f_{e\tau}}{f_{\mu\tau}}\right|
   &\simeq& \frac{\sqrt{2} \Delta m_{\rm atm}^2}{\Delta m_\odot^2}
   \simeq 3\times 10^4
   \label{hi-1}   \,,
 \end{eqnarray}
 from Eq.~(\ref{Zee_mass}) and explicit values of $m_\mu$, $m_\tau$,
 $\Delta m_{\rm atm}^2$ and $\Delta m_\odot^2$.  Here we take the LOW
 solution to the solar neutrino problem \footnote{ As for the other
 solar neutrino solutions, we will give comments in the last section.}.
 These relations induce the ratio of $f_{e \mu}, f_{e\tau}$ and $ f_{\mu
 \tau}$ as
 \begin{eqnarray}
  |f_{e\mu}| : |f_{e\tau}| : |f_{\mu\tau}| =
   1 : 3\times 10^{-3} : 10^{-7}
   \label{hi-2}\,.
 \end{eqnarray}

The phenomenological constraints on $f_{\alpha \beta}$ from various
 experimental bounds~\cite{ST,MN,Mi-Sa} are given by
 \begin{eqnarray}
  &&\frac{|f_{e\mu}|^2}{\overline{M}^2}<3\times 10^{-3}G_F
   \,\,\longrightarrow\,\, |f_{e\mu}| <
   2 \times 10^{-1} \left(\frac{\overline{M}}{1\mathrm{TeV}}\right) ,
   \label{f_{emu}}\\
  &&\frac{|f_{\mu \tau}|^2}{\overline{M}^2},\;\;
   \frac{|f_{e \tau}|^2}{\overline{M}^2}
   <8.5\times 10^{-3}G_F 
   \,\,\longrightarrow\,\, |f_{\mu \tau}| , |f_{e \tau}| <
   3 \times 10^{-1} \left(\frac{\overline{M}}{1\mathrm{TeV}}\right) ,
   \label{other}
 \end{eqnarray}
 where $G_F$ is Fermi constant and
 \begin{eqnarray}
  \frac{1}{\overline{M}^{\;2}}
   \equiv \frac{\sin^2 \chi}{m_{S_1}^2}+\frac{\cos^2 \chi}{m_{S_2}^2}\;.
   \label{barM}
 \end{eqnarray}
 Equations (\ref{f_{emu}}) and (\ref{other}) mean that $f_{\alpha
 \beta}$ cannot be of $\mathcal{O}(1)$ unless the charged Higgs boson
 masses are of order 10 TeV. Notice that the constraints for
 $f_{\mu \tau}$ and $f_{e \tau}$ in Eq.~(\ref{other}) are automatically
 satisfied as long as the relation in Eq.~(\ref{hi-2}) and the
 constraint for $f_{e\mu}$ in Eq.~(\ref{f_{emu}}) are
 satisfied. Actually, the conditions Eqs.~(\ref{hi-2}) and
 (\ref{f_{emu}}) give rise to the following constraint on the smallest
 coupling $f_{\mu \tau}$;
 \begin{eqnarray}
  \label{small-fmutau}
   |f_{\mu\tau}| <
   2\times 10^{-8}
   \times
   \left(\frac{\overline{M}}{1\mathrm{TeV}}\right)
   \,,
 \end{eqnarray}
 which is much severer constraint than that of Eq.~(\ref{other}).

\vspace{5mm}
\section{Cosmological Constraint on the Zee model}
Now let us turn to discuss the cosmological constraint on the Zee model. 
A crucial point is that a linear combination of the lepton flavors,
$L'\equiv L_e-L_\mu-L_\tau$, is approximately conserved in the Zee
model. This is because the $L'$-number is violated only through the
coupling $f_{\mu\tau}$, which is much smaller than the other couplings,
as can be seen in Eq.~(\ref{hi-2}). (Here, we assign the $L'$ number of
the Zee singlet $\omega$ to be zero.) Notice that the sphaleron
preserves not only the total $B-L = B - (L_e + L_\mu + L_\tau)$ but each
$B/3 - L_{\alpha}$ ($\alpha = e,\mu,\tau$). Thus, the number $B/3 + L'$
is also conserved under the sphaleron process, and is violated only by
the tiny coupling $f_{\mu\tau}$.

When the temperature $T$ is higher than the mass of $\omega$,
$M_\omega$, the relevant interactions which violate $L'$ are; $l_\mu
l_\tau \longleftrightarrow l_e l_\mu$, $l_\mu l_\tau \longleftrightarrow
\phi_1 \phi_2$, and $l_\mu l_\tau \longleftrightarrow \omega$. We
consider the three-body process $l_\mu l_\tau \longleftrightarrow
\omega$, since it gives the severest constraint on the coupling
$f_{\mu\tau}$.
\footnote{
If $\mu>M_\omega$, the four-body process
$l_\mu l_\tau \longleftrightarrow \phi_1 \phi_2$ can give a severer
constraint on $f_{\mu\tau}$ than that of the above three-body process.
In this case,
we must take care the possibility
that
the condition $\mu \gg M_{\omega},M_{\Phi}$
might make 
the physical Higgs masses, $m_{S_{1,2}}^2$, be negative.
If the condition $\mu>M_\omega$ and $m_{S_{1,2}}^{2}>0$ is satisfied, we
can analyze the out-of-equilibrium condition of this four-body process
in a similar way to the three-body case.
However, the following conclusion does not change, since the
coupling $f_{\mu\tau}$ is suppressed by $(M_{\omega} / \mu)^2$ for
$\mu > M_{\omega}$ [See Eq.~(\ref{f_mutau})] and hence the
out-of-equilibrium condition is satisfied more easily than the
three-body case.
}
The rate of this process is given by
\begin{eqnarray}
 \Gamma (l_\mu l_\tau \longleftrightarrow \omega)
  &\simeq& 
  \frac{1}{2\pi}|f_{\mu\tau}|^2 M_\omega 
  \left(\frac{M_\omega}{T}\right)
  \,.
\label{rate}
\end{eqnarray}
This $L'$-violating interaction is out of equilibrium if the above
rates is slower than the Hubble parameter of the expanding universe,
$H\simeq (g_* \pi^2/90)^{1/2} T^2/M_\mathrm{PL}$.  ($M_\mathrm{PL}
\simeq 2\times 10^{18}$ GeV is the reduced Planck scale and $g_* \simeq
100$ is the number of relativistic degrees of freedom.) Namely, the
out-of-equilibrium condition is given by
\begin{eqnarray}
 \label{OOE-condition}
 |f_{\mu\tau}|^2
  &\lsim&
  1\times 10^{-14}\left(\frac{M_\omega}{1\,\mathrm{TeV}}\right)
  \,.
\end{eqnarray}

On the other hand, as discussed in the previous section, the coupling
$f_{\mu\tau}$ is very small in order to explain the neutrino oscillation
experiments.  Hereafter, we assume $M_{\omega} \gg M_{\Phi}$ for
simplicity.  In this case, we could consider $m_{S_2} \simeq M_{\omega}
\gg m_{S_1} \simeq M_{\Phi}$. Then, from
Eqs.~(\ref{Zee_mass}),(\ref{mass_atm}), and (\ref{mass_sol}), the
coupling $f_{\mu\tau}$ is given by
\begin{eqnarray}
 |f_{\mu\tau}|^2
  &\simeq&
  64\pi^4
  \frac{M_{\omega}^4}{m_{\tau}^4 \mu^2}
  \frac{(\Delta m_\odot^2)^2}{\Delta m_{\rm atm}^2}
  \nonumber \\
 &\sim&
  10^{-19}
  \left(\frac{M_{\omega}}{1\, \mathrm{TeV}}\right)^4
  \left(\frac{100\, \mathrm{GeV}}{\mu}\right)^2
  \left(\frac{\Delta m_\odot^2}{10^{-7}\, \mathrm{eV^2}}\right)^2
  \left(\frac{3\times 10^{-3}\, \mathrm{eV^2}}{\Delta m_{\rm atm}^2}\right)
\label{f_mutau}\,,
\end{eqnarray}
where we take $\tan \beta =O(1)$ and
$\ln (M^{\;2}_{\Phi}/M^{\;2}_{\omega}) = \mathcal{O}(1)$.

 Here we take the explicit values of $\mu$ and $M_{\omega}$, for
  example, as $\mu\sim \mathrm{100GeV}$ and $M_{\omega} \sim \mathrm{1
  TeV}$. In this case Eqs.~(\ref{OOE-condition}) and (\ref{f_mutau})
  show that the $L'$-violating interactions are really out of
  equilibrium during $T>M_\omega$. This means that the baryon asymmetry
  is {\it not} washed out in the case of LOW solution. This result is
  not changed, unless $M_\omega$ is much heavier than
  $\mathcal{O}(\mathrm{10\,TeV})$ or $\mu$ is extremely small.
 
At the lower temperature $T < M_\omega$, the rate of the three-body
process is reduced because the number density of the $\omega$ particle
is suppressed by a Boltzmann factor $e^{-M_\omega / T}$, and the
four-body interactions are also suppressed by a factor $(T/M_\omega)^4$.

As we have seen in this section, the $L'$-violating interactions have
 been out of equilibrium since the birth of our universe.  This is
 because $f_{\mu\tau}$ must be small in order to explain the neutrino
 oscillation experiments.  Actually, the analysis of chemical potentials
 in the high-temperature phase gives the following relation between the
 baryon asymmetry $B$ and the number $B/3+L'$;
 \begin{eqnarray}
  B=\frac{60}{563}\left({B}/{3}+L'\right)
   \,.
 \end{eqnarray}
 Therefore the baryon asymmetry in our universe can be preserved, once
 the asymmetry of $B/3+L'$ is produced in the early universe.

\vspace{5mm}
 
\section{Summary and Discussion}

Recent neutrino experiments indicate that neutrino masses are tiny and
 there are two mass squared differences as $\Delta m_{\odot}^2 < \Delta
 m_{\rm atm}^2$.  It is well known that the Zee model induces small
 neutrino masses by radiative corrections, where the lepton number is
 violated explicitly.  If the lepton number violating interaction
 becomes faster than the Hubble parameter before the electroweak phase
 transition, the baryon asymmetry generated at higher temperature (in
 the early universe) would be washed out. Therefore, in order not to
 destroy the baryon asymmetry, the lepton number violating interaction
 must be out-of-equilibrium.

In this letter we have analyzed the cosmological condition in order for
 the baryon asymmetry generated in the early universe not to be washed
 out in the Zee model, taking the LOW solution to the solar neutrino
 problem.  In this case, we find the baryon asymmetry is {\it not}
 washed out, although it has been said that the Zee model cannot
 preserve the baryon asymmetry~\cite{Sarkar}. This can be easily shown
 from the view point of the new lepton number $L' \equiv
 L_e-L_\mu-L_\tau$.
 
The lepton number $L'$ is almost conserved quantity which is violated
 only by the tiny coupling $f_{\mu \tau}$. {}From the viewpoint of $L'$,
 the sphaleron process preserves $B/3 + L'$. The $L'$ violating
 interaction through $f_{\mu\tau}$ is out-of-equilibrium in the Zee
 model when the results from neutrino-oscillation experiments and
 natural mass scales of Higgs masses are used for the input
 parameters. Therefore, we can conclude that once the asymmetry of $B/3
 + (L_e-L_\mu-L_\tau)$ is produced, the baryon asymmetry is preserved in
 our universe.

We have shown that the baryon asymmetry is preserved against the
lepton-number violating interaction in the Zee model. However, it is
very difficult to explain the origin of the asymmetry of $B/3 + L'$
within the Zee model. Therefore, some mechanism which can generate a
$B/3 + L'$ asymmetry is necessary.

Finally, we comment on the cases of other solar neutrino solutions (LMA
and VO). First, we consider the LMA case. It has been pointed
out~\cite{noLMA} that the Zee model could not explain the LMA solution,
since the Zee model induces a nearly maximal mixing of solar-neutrino
oscillation ($\sin^2 2 \theta_\odot \simeq 1$), which is in poor
agreement with the observed data~\cite{post-SNO-analysis}. If one would
still apply the present analysis to the LMA solution, the
out-of-equilibrium condition of $L'$ violating interaction is marginally
satisfied [See Eqs.~(\ref{OOE-condition}) and (\ref{f_mutau})], although
it requires a more detailed analysis including a numerical solution of
the Boltzmann equation. The second case is the VO solution. As can be
seen in Eq.~(\ref{f_mutau}), the coupling $f_{\mu\tau}$ which violates
$L'$ is proportional to the $\Delta m_{\odot}^2$. Therefore, in the case
of the VO solution, where $\Delta m_{\odot}^2$ is much smaller than that
of the LOW solution, the rate of the $L'$-violating interaction is also
small enough, and the baryon asymmetry is preserved as discussed in
previous section.

\vspace{5mm}
\section*{Acknowledgment}
We would like to thank T.~Yanagida for the suggestion of this work.  We
 also thank T.~Moroi for the collaboration in the early stage of this
 work, and thank M.~Tanimoto, Y.~Koide 
 and S.~Kanemura for useful discussions.  
NH is supported in part in part by the Grant-in-Aid for Science Research,
 Ministry of Education, Science and Culture, Japan (No.  12740146 ).  
The work of KH was supported by the Japanese Society for
 the Promotion of Science.

%
%



\begin{thebibliography}{99}

\bibitem{Kamiokande}
Y.~Fukuda {\it et al.}  [Kamiokande Collaboration],
Phys.\ Rev.\ Lett.\ {\bf 77} (1996) 1683.

\bibitem {SKatm}
Y.~Fukuda {\it et al.}  [Kamiokande Collaboration],
Phys.\ Lett.\ B {\bf 335} (1994) 237;
\\
Y.~Fukuda {\it et al.}  [Super-Kamiokande Collaboration],
Phys.\ Rev.\ Lett.\ {\bf 81} (1998) 1562
[hep-ex/9807003];
\\
T.~Kajita  [Super-Kamiokande Collaboration], 
in {\it Neutrino Physics and Astrophysics},
Proceedings of the XVIIIth International Conference on Neutrino
Physics and Astrophysics (Neutrino '98), June 4-9, 1998, Takayama,
Japan, edited by Y. Suzuki and Y. Totsuka,
(Elsevier Science B.V., Amsterdam, 1999) page 123;
Nucl.\ Phys.\ Proc.\ Suppl.\ {\bf 77}, 123 (1999)
[hep-ex/9810001].

\bibitem{SKsolar}
Y.~Fukuda {\it et al.}  [Super-Kamiokande Collaboration],
\\
Phys.\ Rev.\ Lett.\ {\bf 81} (1998) 1158
[hep-ex/9805021];
Erratum-ibid.\ {\bf 81} (1998) 4279;
\\
Phys.\ Rev.\ Lett.\ {\bf 82} (1999) 2430
[hep-ex/9812011];
\\
Phys.\ Rev.\ Lett.\ {\bf 82} (1999) 1810
[hep-ex/9812009].

\bibitem{SNO}
Q.~R.~Ahmad {\it et al.}  [SNO Collaboration],
nucl-ex/0106015.


\bibitem{Cl-Homestake}
K.~Lande {\it et al.},
Astrophys.\ J.\  {\bf 496} (1998) 505.


\bibitem{Ga-Gallex-and-GNO}
V.~N.~Gavrin  [SAGE Collaboration],
Nucl.\ Phys.\ Proc.\ Suppl.\  {\bf 91} (2001) 36;
\\
E.~Bellotti,
Nucl.\ Phys.\ Proc.\ Suppl.\  {\bf 91} (2001) 44.


\bibitem{post-SNO-analysis}
V.~Barger, D.~Marfatia and K.~Whisnant,
hep-ph/0106207.
\\
G.~L.~Fogli, E.~Lisi, D.~Montanino and A.~Palazzo,
hep-ph/0106247;
\\
J.~N.~Bahcall, M.~C.~Gonzalez-Garcia and C.~Pena-Garay,
hep-ph/0106258;
\\
A.~Bandyopadhyay, S.~Choubey, S.~Goswami and K.~Kar,
hep-ph/0106264.


\bibitem{seesaw}
T.~Yanagida,
{\it ``Horizontal Symmetry And Masses Of Neutrinos''},
Prog.\ Theor.\ Phys.\  {\bf 64} (1980) 1103,
and in Proceedings of the 
{\it ``Workshop on the Unified Theory and the Baryon Number in the
  Universe''}, Tsukuba, Japan, Feb 13-14, 1979, 
  Eds. O.~Sawada and A.~Sugamoto, KEK report KEK-79-18, p. 95;
\\
M.~Gell-Mann, P.~Ramond and R.~Slansky,
{\it in} ``Supergravity''
(North-Holland, Amsterdam, 1979)
{\it eds.} D.Z. Freedman and P. van Nieuwenhuizen,
Print-80-0576 (CERN).


\bibitem{zee}
A.~Zee,
Phys.\ Lett.\ B {\bf 93} (1980) 389
[Erratum-ibid.\ B {\bf 95} (1980) 461];
%
Phys.\ Lett.\ B {\bf 161} (1985) 141.


\bibitem{sphaleron}
V.~A.~Kuzmin, V.~A.~Rubakov and M.~E.~Shaposhnikov,
Phys.\ Lett.\ B {\bf 155} (1985) 36.


\bibitem{Sarkar}
E.~Ma, M.~Raidal and U.~Sarkar,
Phys.\ Lett.\ B {\bf 460} (1999) 359
[hep-ph/9901406];
%
U.~Sarkar,
Pramana{\bf 54} (2000) 101
[hep-ph/9906335].


\bibitem{bimaximal}
C.~Jarlskog, M.~Matsuda, S.~Skadhauge and M.~Tanimoto,
Phys.\ Lett.\ B {\bf 449} (1999) 240
[hep-ph/9812282],
%
hep-ph/0005147.


\bibitem{ST}
A.~Y.~Smirnov and M.~Tanimoto,
Phys.\ Rev.\ D {\bf 55} (1997) 1665
[hep-ph/9604370].


\bibitem{noLMA}
P.~H.~Frampton and S.~L.~Glashow,
Phys.\ Lett.\ B {\bf 461} (1999) 95
[hep-ph/9906375];
\\
Y.~Koide,
hep-ph/0104226.



\bibitem{Higgs-pheno}
S.~Kanemura, T.~Kasai, G.~Lin, Y.~Okada, J.~Tseng and C.~P.~Yuan,
hep-ph/0011357.


\bibitem{Zee-recent}
D.~Chang and A.~Zee,
Phys.\ Rev.\ D {\bf 61}, 071303 (2000)
[hep-ph/9912380].


\bibitem{MN}
G.~C.~McLaughlin and J.~N.~Ng,
Phys.\ Lett.\ B {\bf 455} (1999) 224
[hep-ph/9903509].


\bibitem{Mi-Sa}
E.~Mituda and K.~Sasaki,
hep-ph/0103202.




\end{thebibliography}
\end{document}